\documentclass[12pt]{iopart}

\usepackage{graphicx}
\usepackage{latexsym}
\usepackage{amsthm}
\usepackage{amscd,amssymb}
\usepackage{floatflt}
\usepackage{mathrsfs}

\begin{document}

\title{A Kirchhoff-like
  conservation law in Regge calculus}


\author{ Adrian P. Gentle${}^{1}$, Arkady Kheyfets${}^{2}$, Jonathan
  R. McDonald ${}^{3}$ \& Warner A. Miller${}^{3}$}

\address{$^{1}$Department of Mathematics, University of Southern Indiana,  
  Evansville, IN  47712 \\
$^{2}$Department of Mathematics, North Carolina State University,  
  Raleigh, NC  27695 \\
  $^{3}$Department of Physics, Florida Atlantic University, Boca
  Raton, FL 33431}

\ead{mcdonald@physics.fau.edu}

\begin{abstract}

  Simplicial lattices provide an elegant framework for discrete
  spacetimes. The inherent orthogonality between a simplicial lattice
  and its circumcentric dual yields an austere representation of spacetime which
  provides a conceptually simple form of Einstein's geometric theory
  of gravitation. A sufficient understanding of simplicial spacetimes
  has been demonstrated in the literature for spacetimes devoid of all
  non-gravitational sources.  However, this understanding has not been
  adequately extended to non-vacuum spacetime models.  Consequently, a
  deep understanding of the diffeomorphic structure of the discrete
  theory is lacking. Conservation laws and symmetry properties are
  attractive starting points for coupling matter with the lattice. We
  present a simplicial form of the contracted Bianchi identity which
  is based on the E. Cartan moment of rotation operator. This
  identity manifest itself in the conceptually-simple form of a
  Kirchhoff-like conservation law. This conservation law enables one
  to extend Regge Calculus to non-vacuum spacetimes and provides a
  deeper understanding of the simplicial diffeomorphism group.
\end{abstract}

\submitto{\CQG}

\maketitle
\setcounter{footnote}{0}

\section{Introduction}

Regge Calculus (RC) provides a natural framework for the description
of discrete spacetimes \cite{Regge:1961,MTW}. It has been applied to
problems ranging from quantum gravity to classical relativity
\cite{RW:1992}. In recent years RC has been used in quantum gravity to
construct emergent spacetime models \cite{Lloyd:2005} and to
investigate the spin-foam approach and its perturbative regime
\cite{Barrett:1999, Bianchi:2007, Freidel:2008, Oriti:2008}.  In particular, 
\cite{Lloyd:2005} utilizes a pre-geometric framework wherein the
spacetime is an emergent structure from an underlying quantum
system. The unitary operators of the discrete quantum system act as
source terms for the ensuing spacetime.  If an approach of this kind
is to utilize RC as the coupling to the underlying quantum dynamics,
it is incumbent upon us to understand the nature of the coupling of
matter and fields to a RC lattice spacetime.

Most early attempts to include matter and fields in RC concentrated on
spacetimes with a high degree of symmetry \cite{Collins:1973,  Dubal:1989} or in specific cases \cite{Sorkin:1975, Hamber:1994}.  A
more general approach was given by \cite{Ren:1988}; however, this
approach introduced a simplex-based action which does not appear to be
natural to us given the vertex-based description indicated by the
contracted Bianchi identity.  A general theory of the coupling of
source terms to RC has not been fully developed. It is the goal of
this manuscript to provide the first steps to construct a simplicial
representation for the stress-energy of the non-gravitational
sources. To do so we first examine the extent to which a simplicial
lattice can guarantee the conservation of energy-momentum. We then
develop more fully the contracted Bianchi identity on the lattice.
This identity, when augmented by the Einstein equations, yields the
conservation of energy and momentum.

We know no shorter route to derive the contracted Bianchi identity
than by using the topological tautology that the boundary of a boundary
is zero, i.e. the boundary-of-a-boundary principle (BBP).  The BBP
appears twice over in each of nature's four fundamental interactions,
once in its 1-2-3-dimensional form, and once in its 2-3-4-dimensional
form \cite{Wheeler:1982b,MTW}.

The BBP has been used in RC to obtain a discrete version of the
contracted Bianchi identity
\cite{Miller:1986,Kheyfets:1989,Kheyfets:1990,Hamber:2004}.  However,
the interpretation of this conservation law has been a source of some
debate particularly over the exactness of the identity.  While the topological principle itself is exact and thoroughly studied in RC \cite{Hamber:2004}, the transition from a continuum
to a discrete spacetime forces one to apply these topological
identity to non-infinitesimal rotations. Unlike the infinitesimal
rotation operators in the continuum, finite rotations do not
ordinarily commute. The transition from the continuum to the discrete
case must be handled with care.  We emphasize that the derivation
presented here will not ordinarily produce an exact identity due to
the non-commuting nature of finite rotations.  Nevertheless what one
loses in exactness one gains in simplicity.  In particular, the
integrated Einstein tensor is doubly projected along its edge and this
allows one to write down the contracted Bianchi identity as a
Kirchhoff-like conservation principle.  These identity are
second-order convergent \cite{Miller:1986} and valid locally at any
event in a spacetime.  

The contracted Bianchi identity for RC has clear implications for
the coupling of energy-momentum to the lattice as well as to our
understanding of diffeomorphism invariance in RC.  
Furthermore, if we expect the quantization of spacetime to produce an
inherently discrete spacetime, then grasping the meaning of the BBP in
a discrete theory becomes essential to understanding the quantum
theory of gravity.  RC serves naturally as an underlying framework
since simplicial spacetimes provide one of the most elegant and
universal descriptions of discrete spacetime \cite{Regge:2000}.

In Sec.~\ref{BBP} we review the BBP and its role in the fundamental
forces of nature.  The importance of this identity stems from its
purely topological foundation.  The Cartan construction of the moment
of rotation trivector in RC is reviewed before applying the BBP
directly to the simplicial lattice in Sec.~\ref{BBPRC}.  We conclude
in Sec.~\ref{Discussion} with our future plans to couple a generic
stress-energy tensor to the geometric content of the Regge lattice.

\section{Boundary of a Boundary Principle: The Guiding Topological
  Principle} 
\label{BBP} 

In any fundamental field theory (electrodynamics, Yang -- Mills,
general relativity) the conservation of source is introduced in such a
way that it is satisfied for any field.  This is equivalent to saying
that it does not impose any restrictions on the field itself but
rather puts constraints on the source of the field (charge in
electrodynamics, energy--momentum in general relativity). This feature
is conditioned only by the requirement that the field is described as
the curvature of a connection on the appropriate vector bundle that is
responsible for the correct implementation of the field symmetries
\cite{Bleecker}.

\begin{figure}[htp]
\centering 

\includegraphics[height=3in]{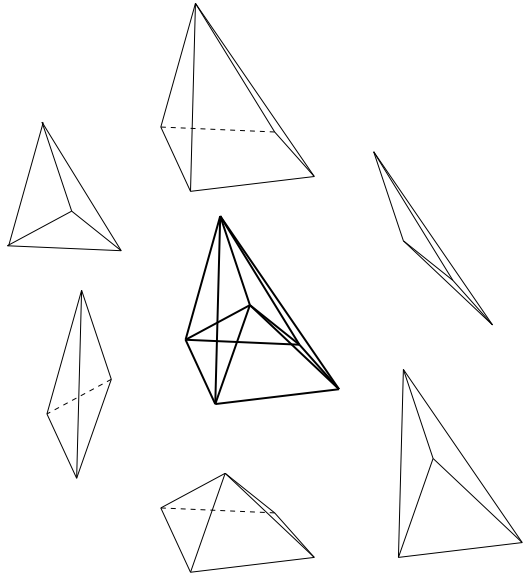}

\caption{{\emph{The polyhedral boundary of a 4-polytope:}} This
  illustration shows the 2-dimensional projection of a typical
  4-dimensional polytope, $V^*$, of the circumcentric dual (Voronoi)
  spacetime. It is dual to a vertex, $V$, of the simplicial (Delaunay)
  spacetime. This 4-polytope is bounded by six polygons (shown
  exploded off into the perimeter of the polytope). These 4-polytopes
  are ordinarily not  simplexes nor are their bounding polyhedra.
  The orientation of $V^*$ induces an orientation on each of its
  polyhedral faces, $L^*$.  The orientation of each polyhedron
  consequently induces an orientation on each its polygonal
  faces. However, each 2-face is shared by two polyhedra thereby
  inducing equal and opposite orientations on it.  In this sense none
  of these polygonal faces are exposed and their orientations
  cancel. This is the origin of the BBP principle in its 2-3-4
  dimensional form.  }
\label{fig:4V}
\end{figure}

The universality of this feature follows from the fact that it is
induced by (but not totally reduced to) the simple topological
identity that the boundary-of-a-boundary is equal to zero
\cite{Cartan:1928}.  Application of this principle to spacetime is
achieved by associating to it a chain complex (say by simplicial or
any other triangulation) with the standard boundary operator based on
the rules of orientations of the boundaries. As an example, we can
examine a discrete representation spacetime wherein the spacetime
geometry is tiled by 4-dimensional polytopes. The geometry interior to
each of these infinitesimal polytopes is irrelevant and, for pictorial
representations, can be thought of as flat Minkowski geometry.  Let us
examine one of these polytopes, $V^*$, which is the local neighborhood
of an event, $V$.  This polytope is bounded by 3-dimensional polyhedra
(Figure~\ref{fig:4V}).  Any two adjacent polyhedra on the boundary of
$V^*$ share a common 2-dimensional face.  In other words, in this
4-dimensional region of spacetime no 2-dimensional polygonal faces
are exposed. In general relativity, any flow of stress-energy (or
equivalently the dual of the Cartan moment of rotation) into one of
the 3-dimensional bounding polyhedra is exactly compensated by an
equal flow of stress-energy (Cartan moment of rotation) out of an
adjacent polyhedron. This guarantees conservation of source in $V^*$.

\begin{figure}[htp]
\centering 

\includegraphics[width=3in]{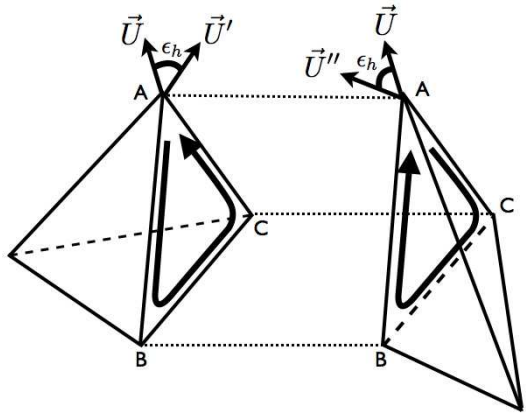}

\caption{{\emph{BBP as a Geometric Identity:}} Here two adjoining
  faces of the 3-dimensional boundary of a 4-dimensional volume are
  depicted with their induced orientation.  The orientation of the
  2-dimensional area is seen to be opposite for the adjoining
  3-volumes such that in the sum over the boundary of the boundary
  these areas cancel one another.  Furthermore, if a vector $\vec{U}$
  is parallel transported around the area adjoining the two 3-volumes,
  then the vector will ordinarily come back rotated.  When the area is
  associated with the left 3-volume, the vector $\vec{U}$ comes back
  rotated as $\vec{U}^{'}$, but when the area is associated with the
  right 3-volume it will come back as $\vec{U}^{''}$.  The rotation in
  both cases is in the same plane and rotated by the same amount but
  in opposite directions of rotation.  } \label{fig:bbp}

\end{figure}

For each $V^*$, one would like to sum over each of its unexposed
2-dimensional boundaries -- the meeting place of two of the polyhedral
boundaries of $V^*$ (Figure~\ref{fig:bbp}). These polyhedral
boundaries induce opposing orientations on each of the 2-dimensional
faces.  Therefore when one sums over all of the 2-boundaries of all
the 3-boundaries, two contributions are found for each polygon each of
equal magnitude but opposite orientations. These identically cancel
one another leaving the boundary-of-a-boundary identically equal to
zero.

In applications to continuum field theories the boundary of a boundary
relation of the chain complex is translated into the relation
co-boundary of a co-boundary of the dual de Rham co-chain complex of
differential forms.  The exterior derivative acts as the co-boundary
operator \cite{Cartan:1928}. This duality is established by adding
rotations caused by parallel translations of vectors around the
2-faces (or Cartan moments of these rotations) of an infinitesimal
3-simplex (or 4-simplex for moments of rotations).  These rotations
are expressed as products of Riemannian curvature tensors on each face
and the oriented element of area associated with the face. This
operation, when applied to finite structures, is ambiguous and cannot
be performed in a consistent way.  The ambiguity arises for two
reasons: finite rotations do not commute, and tensor quantities are
being computed at different points (on different faces) and then
added.  These difficulties disappear in the infinitesimal limit.

In particular the application of 1-2-3 dimensional
BBP in general relativity reduces to computing the sum of rotations caused by parallel
transport around all 2-faces of a 3-simplex. When expanded in Taylor
series with respect to displacements along the edges \cite{MTW} it
produces terms of second and the third order (the higher orders are of
no interest in this computation).  The ambiguity caused by parallel
transporting tensor quantities to a common point (necessary for
addition) introduces errors of the fourth and higher order.  These
errors can be neglected. The requirement that the second order term
vanishes leads to the conclusion that the Riemannian curvatures on the
faces of the 3-simplex are not linearly independent, while the
requirement that the third order term vanishes implies the ordinary
Bianchi identity.

Likewise, application of 2-3-4 dimensional BBP amounts to adding
Cartan moments of rotation over all 3-faces of a 4-simplex (or
polytope).  The Taylor series expansion proceeds as before, with terms
up to the third order disappearing because of relations imposed by the
1-2-3 dimensional form of the BBP.  The errors generated by the
ambiguity of parallel translation are now of the fifth and higher
order.  The contracted Bianchi identity arises from the fourth order
term of the expansion.

\section{Discrete Bianchi Identity} 
\label{BBPRC} 

Regge Calculus is based on a lattice of flat 4-dimensional simplexes
that form a curved piece-wise flat manifold.  The curvature is
concentrated as conical singularities on each of the co-dimension 2
triangular hinges.  To define the curvature we need to associate an
``area-of-circumnavigation'' to each triangular hinge. This provides a
finite area over which we can distribute the curvature.  Fortunately,
the circumcentric dual lattice has been shown to arise naturally in RC
and provides an appropriate area
\cite{CFL82,CFL82b,L83,FL84,FFLR84,CDM89,HW86,HW86b,M97,McDonald:2008a,McDonald:2008b}.
Correspondingly, it has been shown that the circumcentric 3-volumes
naturally define the moment-of-rotation operators and discrete RC
equations \cite{Miller:1986}. Finally, we postulate here that the
circumcentric 4-volume ($V^*$) dual to a vertex ($V$) defines a
natural domain to apply the Cartan BBP in its 2-3-4-dimensional form.
Consequently the BBP in RC becomes the ``co-boundary of the
co-boundary'' principle, although the geometric underpinnings are
exactly the same.

In this section we derive the discrete form of the contracted Bianchi
identity.  We begin by emphasizing the central role that the Cartan
moment-of-rotation trivector and the circumcentric dual lattice play
in this derivation. In particular, we begin by re-expressing the
familiar Regge equations in the Cartan prescription. This leads
naturally to a Kirchhoff-like identity at each vertex inherently
linked with the topological boundary-of-a-boundary identity. We
conclude the section with an analysis of the convergence properties of
the identity with the typical lattice edge length $L$.

\subsection{The Regge-Einstein tensor and the Cartan moment of
  rotation}  \label{Cartan-RC} 

To begin the derivation we follow E. Cartan and examine the moment of
rotation trivector.  The dual to this trivector generates the Regge
equation as well as the RC version of the Einstein
tensor. Recall that there is one Regge equation for each edge, $L$, in
the simplicial spacetime geometry,
\begin{equation}
\label{RE}
\left(\begin{array}{c} $Regge$\ $equation$ \\
                       $associated\ with$\\
                       $edge$\ L \end{array} \right) = 
\sum_{h\supset L} \frac{1}{2}\, L\, \cot{\left(\theta_{Lh}\right)} \,\epsilon_h,
\end{equation}
where the sum is taken over each triangular hinge $h$ sharing edge
$L$, $\theta_{Lh}$ is the interior angle of $h$ opposite $L$, and
$\epsilon_h$ is the deficit angle of hinge $h$.

The Cartan moment-of-rotation trivector is defined through a moment
arm ($dP$) reaching from a fulcrum to a rotation bivector.  Each
triangular hinge $h$ in the simplicial spacetime has an associated
rotation bivector (${\cal R}_h$) located at the circumcenter $C$ of
the hinge $h$. The orientation of ${\cal R}_h$ is in a 2-plane, $h^*$,
perpendicular to the hinge, $h$.  The bivector is formed by two unit
vectors separated by the usual RC deficit angle,
$\epsilon_h$.

It is convenient to locate the fulcrum at one of the two endpoints
edge $L$.  We denote this fulcrum vertex as $V$, and by construction
it is one of the vertices of hinge $h$.  This freedom of choice is
guaranteed by the ordinary Bianchi identity as we show below.  This
is in contrast to previous derivations of the Regge equations using
the Cartan approach where the fulcrum was taken halfway along edge
$L$ \cite{Miller:1986,McDonald:2008a,McDonald:2008b}.

\begin{figure}[htp]
\centering 

\includegraphics[width=4in]{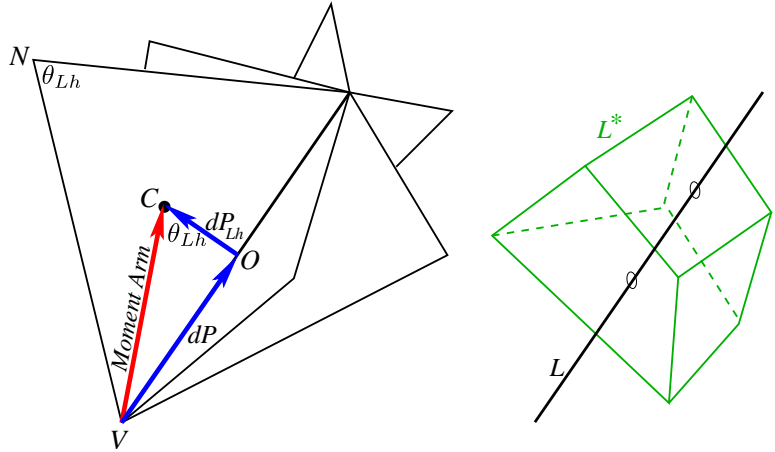}

\caption{{\emph{Hinges and Moment Arm:}} In the simplicial lattice
  each edge is common to multiple hinges $h$ (left).  The
  circumcentric 3-volume $L^*$ dual to edge $L$ has 2-dimensional
  boundaries dual to each of the hinges $h$ (right).  The parallel
  transport of a vector around the perimeter of these dual areas will
  result in a net rotation by an angle equal to the deficit angle,
  $\epsilon_h$, associated with the hinge, $h$.  The moment of
  rotation is given by a moment arm $P_{L}+dP_{Lh} $ wedge the
  rotation associated with the parallel transport around the dual
  area.  However, the first term does not contribute as it is equal to
  zero by the ordinary Bianchi identity. On a given hinge, the
  effective moment arm is the vector from the edge to the center of
  rotation, i.e. the circumcenter of the hinge $C$, which has length
  $(1/2) L \cot{\theta_{Lh}}$.  } \label{fig:hinge}

\end{figure}

With the fulcrum at $V$ we can decompose the moment arm associated
with  hinge $h$ into two vectors (Fig. \ref{fig:hinge}),
\begin{equation}
\label{MA1}
\left( \mathrm{Moment\, Arm}\right)_{Lh}  = {P}_{L} +
d{P}_{Lh} 
\end{equation}
where $P_{L}=\frac{1}{2}\bf{L}$ is the vector from the fulcrum ${V}$
to point ${O}$, located at the center of edge $L$.  This is also the
center of three-dimensional circumcentric polyhedron ${L}^{*}$,
defined to be dual to edge $L$.  The other component of the moment
arm, $d{P}_{Lh}$, is the vector from ${O}$ to the circumcenter ${C}$
of the hinge.  This gives us two vectorial contributions to the moment
arm;  one (${ P}_L$) is common to all 2-dimensional faces $h^*$ of the
dual polyhedron ,$L^*$, and another ($d{P}_{Lh}$) is distinct for each
of these 2-dimensional faces. The contribution common to all faces of
$L^{*}$ can be factored out of the sum of moments of rotations, so
that
\begin{equation}
\label{fulcrummove}
\sum_{h\supset L} \, \left( {\cal P}_L + d{\cal P}_{Lh} \right) \wedge {\cal R}_h
= \underbrace{{\cal P}_L}_{\mathrm{common}} \wedge {\sum_{h\supset L} {\cal R}_h}\  +\  
 \sum_{h\supset L} d{\cal P}_{Lh} \wedge {\cal R}_h  .
\end{equation}
The resulting sum over all rotations around ${L}^{*}$ is simply the
ordinary Bianchi identity for RC \cite{Regge:1961, Miller:1986},
\begin{equation}
\label{obi}
\sum_{h\supset L}\,{\cal R}_h = {\mathcal O} (L^2).
\end{equation}
In Eqs. \ref{fulcrummove}-\ref{obi} the sum over the hinges, $h_L$,
sharing edge $L$ could have equally been taken over the bounding
polygons, $h^*$, of the dual polygon, $V^*_L$.  There is a one-to-one
correspondence between the $h$ and $h^*$.

Using this approximate Bianchi identity we are justified in removing the
common contribution to the moment arm in our sum over the
moments-of-rotation.  We see that the ordinary Bianchi identity allows
us to freely choose the position of the fulcrum.  A natural choice for
the fulcrum is the vertex $V$ and we use ${V}^{*}$ to denote the dual
4-polytope to this vertex.  Then each edge, $L$, emanating from vertex,
${ V}$, has the moment arm, $P_{L}=\frac{1}{2}\bf{L}$, which is directed
along edge, $L$.  Since each edge $L$ is dual to a corresponding
3-polytope, $L^*$, the effective moment arm is
\begin{equation}
\label{MA2}
\left( \begin{array}{c}$Effective$ \\ $Moment Arm$ \end{array} \right) =
dP_{Lh} =  \frac{1}{2}\,L\,\cot{\theta_{Lh}}\, \mathbf{\hat{n}},
\end{equation} 
which is the segment from the center of the edge ($O$) to the
circumcenter ($C$) of the hinge (Figure~\ref{fig:hinge}).

We are now in a position to explicitly reconstruct the Regge equation
associated to an edge $L$, and to construct the corresponding
Regge-Einstein tensor.  To define the moment of rotation trivector
associated to hinge $h$ and edge $L$ we needed both the moment arm and
the rotation bivector.  Parallel transport of a unit vector around the
2-dimensional face $h^*$ dual to the hinge $h$ returns a unit vector
rotated by an amount equal to the deficit angle ($\epsilon_h$)
associated with the hinge.  Furthermore, the rotation bivector lies in
the plane $h^*$, perpendicular to the hinge.  

The dual of the Einstein
tensor is  expressible in terms of the moment of rotation trivector
\cite{MTW},
\begin{equation}
\label{momrot}
  \int_{V^*}{}^{*}\bf{G} = \int_{\partial V^*}  
\star\left( dP \wedge {\cal R}\right) = 0,
\end{equation}
where the Hodge dual only acts in the space of values, i.e. on the
moment of rotation trivector.  In RC the moment of rotation trivector
consists entirely of the parallelepiped formed by the moment arm
$dP_{Lh}$ and the two vectors defining the rotation bivector,
\begin{equation}
\label{rcbivector}
\left( \begin{array}{c} 
$Bivector$ \\
$dual\ to$ \\
$hinge$\ h \end{array} \right) = {\cal R}_h =
\frac{\mathbf{L} \wedge \vec{VN}}{2 {A}_h} \, \epsilon_h,
\end{equation}
which lies in the plane orthogonal to the triangular hinge $h$.  This
hinge is defined by the vectors $\mathbf{L}$ and $\vec{VN}$, and has
area ${A}_h$. The star dual of the parallelepiped returns a vector of
length $\epsilon_h$ and parallel to edge $\mathbf{L}$.

We can now construct the moment-of-rotation trivector. The dual moment
of rotation associated with a hinge $h$ containing the edge $L$ is
\begin{equation}
\label{hingemoment}
\left( \begin{array} {c}  \mathrm{Dual\, Moment} \\
    \mathrm{of\, Rotation} \\
    \mathrm{for\; hinge\; }h \end{array} \right)_{L}
=  {\star}\left(  \mathrm{d}{P}_{Lh} \wedge
  {R}_{h}\right) \quad \longrightarrow \quad
\underbrace{\frac{1}{2}\,\mathbf{L}
  \cot{\theta_{Lh}}}_{\mathrm{Moment\;Arm}}  
\underbrace{\epsilon_{h}}_{\mathrm{Rot'n}}.
\end{equation}
The {\sl total} dual moment of rotation over the Voronoi 3-volume
$L^*$ is then found by adding contributions from all hinges which
share the edge $L$,
\begin{equation}
\label{totalmoment}
\int_{\partial\cal{V}^*} {\star}\left( \mathrm{d}{P}_{Lh}
  \wedge {R}_{h}\right) \quad \longrightarrow \quad
\frac{1}{2}\,\sum_{h\supset L}\, \mathbf{L}\, \cot{\theta_{Lh}}\,
{\epsilon_{h}}.
\end{equation}

In the Cartan description of Einstein's theory \cite{MTW,
  Wheeler:1982b, Cartan:1928} the Einstein tensor associated with a
three dimensional region is the dual of the total moment of rotation
tri-vector per unit three-volume.  The two components of the Einstein
tensor describe the orientation of the three volume, and the
orientation of the moment of the rotation tri-vector.  In Regge
calculus there is one equation per edge $L$, as can be seen when the
Cartan moment of rotation is calculated over the Voronoi three-volume
$L^*$ \cite{Miller:1986}.  The orthogonality between the simplicial
(Delaunay) lattice and its circumcentric dual (Voronoi lattice) yields
an Einstein tensor which is doubly projected along edge $L$.  That is,
\begin{equation}
\label{reggemoment}
  \left( \begin{array} {c}  \mathrm{Integrated\, Einstein} \\
      \mathrm{Tensor\, associated  } \\
      \mathrm{with\; edge\;}L \end{array} \right)
 = \int_{V^*}  {}^\star G \quad  \longrightarrow \quad
\frac{G_{LL}\ L^*}{L} \, \mathbf{L},
\end{equation}
which is directed along edge $L$ and has magnitue $G_{LL} L^*$.

Combining equations (\ref{momrot}), (\ref{totalmoment}) and
(\ref{reggemoment}) establishes the relationship between the Regge
equations and the {\emph{integrated}} simplicial Einstein equations
\cite{Miller:1986},
\begin{equation} 
\label{ERequations} 
G_{LL} \, L^* =
\frac{1}{2} \, \sum_{h\supset L}\, L\, \cot{\theta_{Lh}}\, \epsilon_{h}.
\end{equation}
This effectively defines the simplicial Einstein tensor $G_{LL}$ at
edge $L$.

Finally, we note that the simplicial Einstein tensor along the edge
$L$, constructed using the sum of moments of rotations for the dual
3-volume $L^*$, is simply the geometric portion of the familiar Regge
equation,
\begin{equation}
\frac{1}{2}\, \sum_{h\supset L}\, L\, \cot{\theta_{Lh}}\, \epsilon_{h}.
\end{equation}

\subsection{The contracted Bianchi identity}

As shown above, the Regge equation and corresponding simplicial
Einstein (or Regge-Einstein) tensor is naturally defined relative to
the dual polyhedron $L^*$.  It is therefore natural to define the
4-dimensional polytope ${V}^*$ dual to the vertex ${V}$ as the domain
upon which we apply the Cartan BBP in its 2-3-4 dimensional form. To
demand no net creation of source ($\nabla \cdot G = \nabla \cdot T =
0$) in this spacetime region is to embody the essence of the
contracted Bianchi identity.

As in the continuum (Sec.~\ref{BBP}), we can provide a finite sum
(``integral'') representation of the contracted Bianchi identity
associated to the dual polytope ${ V}^*$ by summing over its
polyhedral 3-boundaries $L^{*}$.  By construction, each polyhedral
3-boundary of the dual polytope ${L}^*$ is dual to one of the edges
$L$ of the simplicial lattice emanating from vertex $V$.  This defines
the domain of integration for the BBP.  This completes two steps
toward the BBP in RC by defining both the domain and the integrand
(Sec. \ref{Cartan-RC}) for the BBP in RC.


The final step in deriving an expression for the conversation of
moment of rotation in RC is achieved by summing over the dual
3-volumes $L^{*}$ that bound the 4-volume $V^*$, which is dual to
vertex $V$.  However, care must be taken in evaluating this sum.
Despite our choice of a common fulcrum, we have still decomposed the
total moment arm into two vectors (one strictly in the tangent space
of the vertex $V$, and one at the center of edge $L$).  Yet, RC
provides a simple solution to what could be problematic.  Both of
these vectors lie in the tangent space of their associated hinge $h$.
As such, the decomposition of the total moment arm becomes the
standard decomposition of a vector in flat Minkowski spacetime.
Summation over terms at a common point can be achieved in two
equivalent, but separate approaches; (1) by parallel transporting the
effective moment arm prior to inclusion in the moment of rotation
trivector, or (2) parallel transporting the net moment of rotation
trivector.  We will consider the second approach.


In RC the integrated Einstein tensor (\ref{ERequations}) is not only
evaluated along the edge $L$, it is also directed along $\mathbf{L}$.
Since each simplicial edge $L$ is by definition a geodesic in the
lattice, any vector parallel transported along $\mathbf{L}$ will
maintain a constant angle with respect to $\mathbf{L}$.  We take
advantage of this property and individually transport each of the
RC moment-of-rotation trivectors from the center of their
respective edges to the vertex $V$ which is common to all of these
edges.  We are then free to sum these moment-of-rotation vectors at
$V$.  Repeating this procedure across the lattice yields a 4-vector
identity at each and every vertex $V$,
\begin{equation}
\label{net_mor}
\left( \begin{array}{c}
$Net\ Moment$ \\
$of\ Rotation$ \\
$at\ vertex$\ V \end{array} \right) =
\sum_{L\supset V} \;\; \; \sum_{h\supset L} \;\;\;   
\frac{1}{2}\, \mathbf{L} \,\cot{\left(\theta_{Lh}\right)} \, \epsilon_{h}.
\end{equation}
   
This is the simplicial form of the net moment of rotation at vertex
$V$ and must vanish by the 2-3-4 dimensional form of the BBP.
However, as we have mentioned, the finite rotation operators do not
ordinarily commute.  This is important because we must apply our
rotations in a given order.   Nevertheless, the non-commutativity of the rotation operators can be
  made as small as one wishes by suitably refining the lengths of the
  simplicial lattice.  Here, suitable refinement of the lattice is
  taken in the sense described in \cite{CMS:1984} where constant
  curvature barycentric subdivision is employed to refine the edge
  lengths by introducing new simplicial blocks and distributing
  curvature over the new subdivision of the simplexes.  Under such
  refinements, the commutators for rotations scale as the deficit
  angles squared.  Moreover, the deficit angles scale as the edge
  length squared as can be seen via their relation to the curvature
  $\left( \mathrm{Curvature}\right)= K = \epsilon_{h} / A^{*}_{h} $.
  Consequently, the deficit angles scale as $\mathcal{O}(L^{2})$ and
  the commutators for rotations scale as $\mathcal{O}(L^{4})$.  This
  second order convergence is the origin of the approximation being
  implemented. Therefore, 
\begin{equation}\label{eq:ERcBI}
  \underbrace{\sum_{L\supset V} \;\; \; \sum_{h\supset L}}_{\delta
    \circ\delta \equiv 0} \;\;\;   \frac{1}{2}\, \mathbf{L}
  \,\cot{\left(\theta_{Lh}\right)} \, \epsilon_{h} +
  {\mathcal O}(L^5) = 0.
\end{equation} 
This is the RC formulation of the contracted Bianchi identity.  The
first term in this expression scales with $\mathcal{O}(L^4)$, since the deficit
angles $\epsilon_h$ scale as $\mathcal{O}(L^2)$.  The contracted Bianchi identity is not identically zero because small, finite, rotations do not
necessarily commute.  Consequently, the final term scales with both
the edge length $L$ and the rotation commutator $[\epsilon_{h},
\epsilon_{h'}]$, yielding an overall $\mathcal{O}(L^5)$ behaviour in the error
term.


Two features are apparent in the discrete contracted Bianchi identity.  First, it has the form of a Kirchhoff-like conservation law.   The analysis presented in this manuscript completes the derivation of this Kirchhoff-like property of the contracted Bianchi identity. We have reduced the results of previous calculations from a non-local, boundary-valued sum to a vertex-based conservation equation. This was accomplished by utilizing the freedom we have in choosing the fulcrum for each of the moment of rotation trivectors, here we have chosen the vertex $V$ common to all of the faces of the dual polytope $V^{*}$.  Equivalently, this can be understood by our ability to parallel transport the moment of rotation trivector from the midpoint $\mathcal{O}_{i}$ the edge $L_{i}$ to the vertex $V$.  No higher order corrections than corrections already discussed in this manuscript are introduced.   It is this understanding of the relation between the contracted Bianchi identity gives rise to conservation of source which is vital to understanding how source can be coupled to the lattice.  Second, the appearance of a 4-vector identity at each vertex signals that there are exactly four ``approximate'' diffeomorphic degrees of freedom per vertex in the simplicial lattice.   This last
point has been important for understanding the dynamical degrees of
freedom in RC \cite{Miller:1986, Kheyfets:1989, Kheyfets:1990}, and
the resulting approximate diffeomorphism freedom has been utilized to
solve the initial value problem \cite{Gentle:1998}.

\begin{figure}[ht]
\centering

\includegraphics[height=2in]{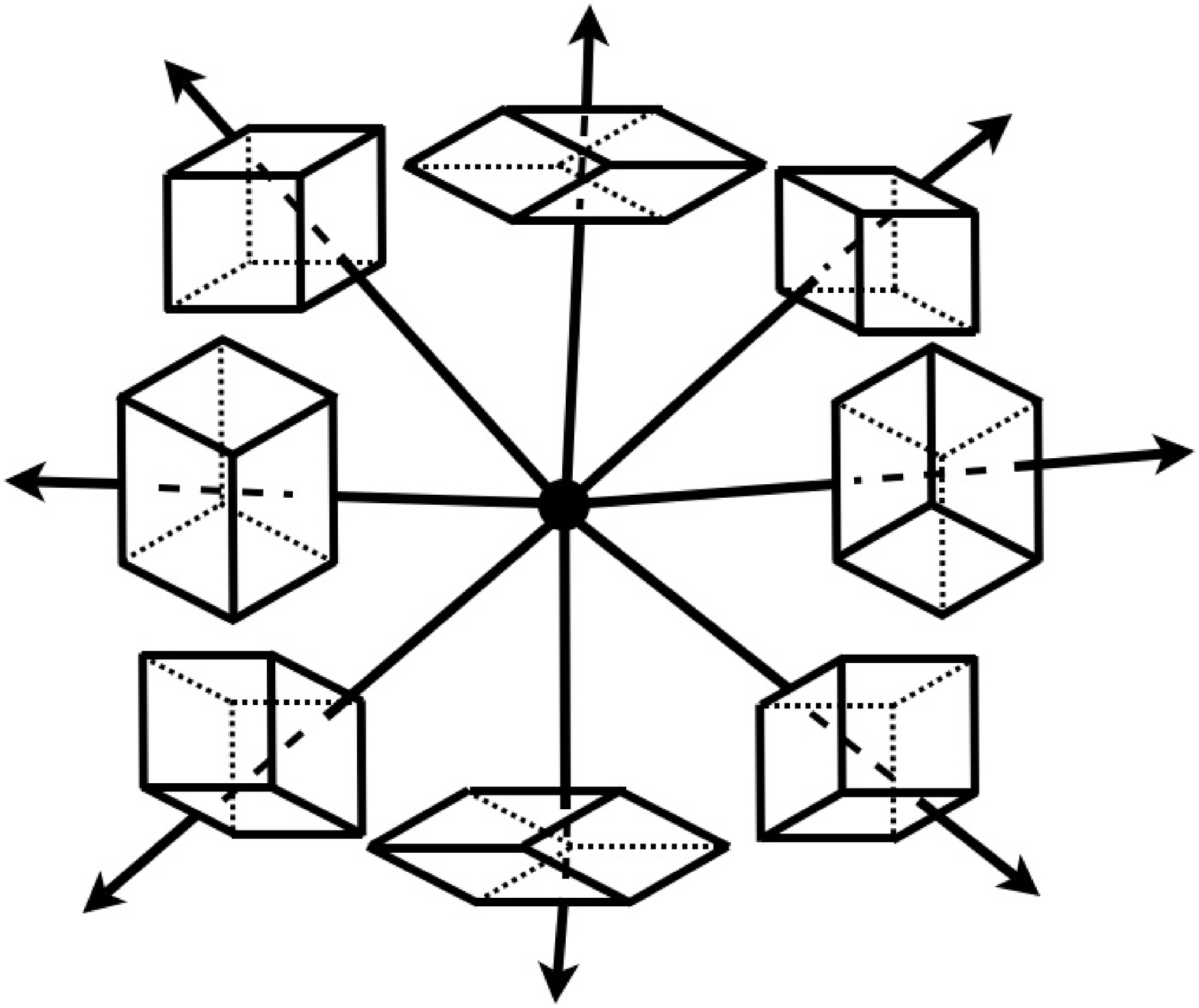}
\includegraphics[height=2in]{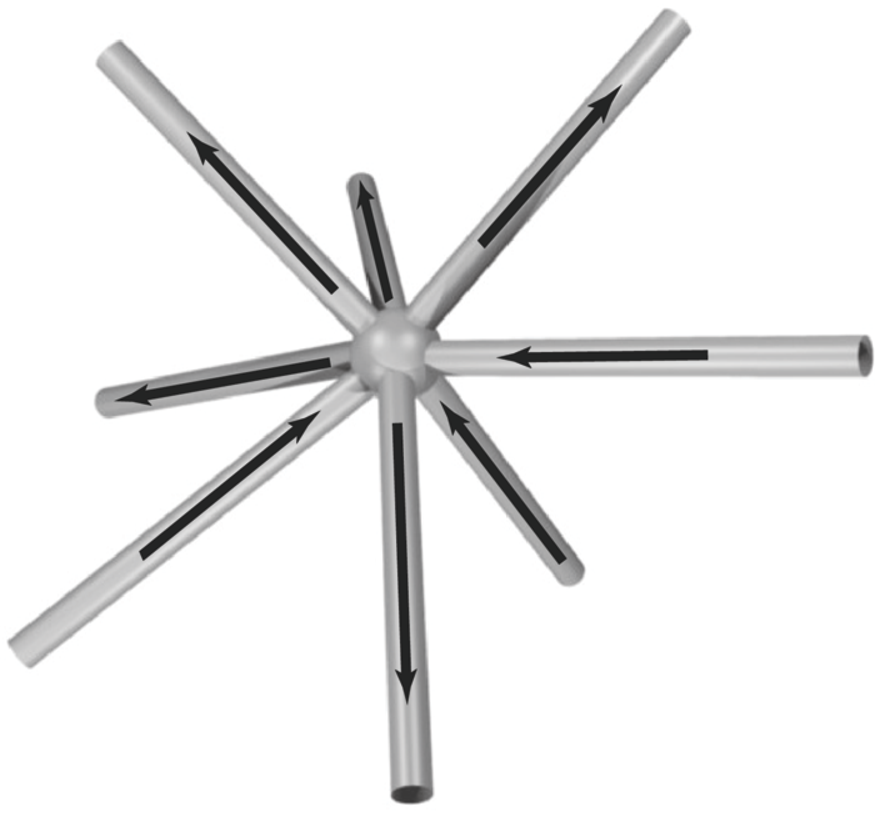}
 
\caption{{\emph{The Kirchhoff-like form of the contracted Bianchi
      identity in RC:}} On the left is an exploded view of the edges
  meeting a vertex $V$ and their dual 3-volumes, $L^*$.  The
  first step towards the Kirchhoff-like conservation principle is
  constructing the total moment of rotation for each of the dual
  3-volumes, $L^*$.  On the right is a depiction of the flow of
  moment-of-rotation, or equivalently flow of stress-energy along each
  of the edges, $L$ meeting at the vertex $V$.  The flow of
  moment-of-rotation entering or leaving vertex, $V$, is
  conserved to second order in the lattice spacing, $L$. Since the
  Einstein equations, and their RC equivalent, equates the
  moment-of-rotation with the stress-energy, this contracted Bianchi
  identity can be viewed as a circuit-like conservation law. Here the
  ``wires'' of the circuit are the edges of the simplicial lattice,
  and the ``current'' in each of the wires is the doubly-projected
  stress-energy tensor $T_{LL}$ along the given edge $\mathbf{L}$
  emanating from vertex $V$.  }
\label{fig:Kirchhoff}

\end{figure}

\section{Expanding Regge Calculus Beyond the Vacuum} 
\label{Discussion} 

RC has been a predominately vacuum theory of gravitation with only
specialized applications to spacetimes with non-gravitational
sources \cite{RW:1992}. Our approach can provide a consistent and
generic coupling of matter to the simplicial lattice. The cornerstone
of our approach relies on the nature of conservation principles of the
lattice.  We know of no better way to understand the conservation than
through the BBP.  This application of the BBP has led to an
approximate conservation of moment-of-rotation for RC.  

The conservation of moment-of-rotation takes a form which is ideally
suited for applications to matter: the contracted Bianchi identity
in RC becomes a Kirchhoff-like conservation principle along the edges
emanating from a specific vertex.  With this and Einstein's equations, we
obtain an approximate conservation equation for energy-momentum on the
lattice,
\begin{equation}
\sum_{L\supset V} G_{LL} L^{*} =\sum_{L\supset V}
\kappa T_{LL} L^{*} \cong 0.
\end{equation}
In particular, the doubly projected stress-energy along the edges
emanating from a vertex must sum to zero, to at least second order in
the length scale of the lattice.  While not exact, this gives the
interpretation of a Kirchhoff-like conservation principle for the
geometry, and (with Einstein's equations) the flow of energy and
momentum (Figure~\ref{fig:Kirchhoff}).  As a result, we obtain a set of vertex-based constraints
for edge-based expressions that constrain energy-momentum.  This
exercise indicates that energy-momentum is naturally wired to the
simplicial lattice at each vertex
and is naturally wired to each edge in its coupling with the
simplicial field equations.

For applications of RC to pre-geometric quantum spacetime, one must
necessarily formulate an appropriate stress-energy tensor arising from
the quantum dynamics. For applications to classical spacetimes a
simplicial form of the stress-energy tensor must be constructed from
the non-gravitational sources.  This work indicates that the
stress-energy will most naturally be expressed as a vertex-based
tensor, and that its coupling to the RC equations will be through its
double projection on the edges of the lattice.  We will explore this
coupling in future work.

\ack 

We were fortunate to have been exposed to the deep insights and motivations of John Archibald Wheeler.  His valuable contributions to this work made it possible to provide such a clear geometric model of conservation of energy-momentum in a discrete spacetime.


\end{document}